\documentclass[a4paper,10pt]{article}
\usepackage{
times,
amsmath,amssymb,epsfig,graphics,graphicx}

\newcommand{\be}{\begin{eqnarray}}
\newcommand{\ee}{\end{eqnarray}}

\begin{document}

\date{\today}

\begin{center}
{\Large \bf  Perturbing free motions on hyper-spheres without degeneracy lift }
\end{center}

\vspace{0.02cm}
\begin{center}
A.\ Pallares-Rivera
\footnote{e-mail: pallares@ifisica.uaslp.mx}, 
F. de J. Rosales-Aldape
\footnote{e-mail: r\_felipedejesus@yahoo.com.mx}, 
M.\ Kirchbach
\footnote{e-mail: mariana@ifisica.uaslp.mx}
\end{center}

\vspace{0.02cm}

\begin{center}
Institute of Physics\\
Autonomous University of San Luis Potos{\'{i}}\\
Av. Manuel Nava 6, University Campus\\
SLP 78290 San Luis Potos{\'{i}}, M\'exico
\end{center}

\begin{quote}{\bf Abstract:}
{We consider quantum motion on $S^3$ perturbed by the trigonometric Scarf potential (Scarf I) 
with one internal quantized dimensionless parameter, $\ell$, the  ordinary orbital angular momentum value,
and another, continuous parameter, $b$, through which an external scale is introduced.
We argue that a loss of the geometric hyper-spherical $so(4)$ symmetry of the free motion 
occurs that leaves intact the unperturbed hydrogen-like degeneracy patterns characterizing the spectrum under discussion.
The argument is based on the observation that the expansions of the Scarf I wave functions for  fixed  $\ell$-values
in the basis of properly identified $so(4)$  representation functions are power series in the perturbation parameter, $b$, 
in which carrier spaces of dimensionality $(K+1)^2$ 
with $K$ varying as $K\in [\ell , N-1]$, and $N$ being  the principal quantum number of the Scarf I potential problem,
contribute up to the order ${\mathcal O}\big({b}^{N-1-K}\big)$. 
Nonetheless, the degeneracy patterns can still be interpreted as a consequence of  an effective $so(4)$ symmetry, i.e. a symmetry
realized at the level of the dynamic of the system, in so far as from the perspective of the eigenvalue problem, 
the Scarf I results are equivalently obtained from a Hamiltonian with matrix elements of polynomials in a 
properly identified $so(4)$ Casimir operator.
The scheme applies to any dimension  $d$.
}
\end{quote}

\begin{flushleft}
\end{flushleft}
\vspace{0.02cm}

\begin{flushleft}
\end{flushleft}


\section{Symmetry and degeneracy: Introductory remarks}

Symmetry and degeneracy are two concepts which one traditionally associates with the basics of the  
quantum mechanics teachings. One may think of the degeneracy with respect to the magnetic quantum number, $m$,
of a quantum level of a given angular momentum $\ell$ which is $(2\ell+1)$-fold, and typical for the states bound within
all central potentials. A more advanced example would be the degeneracy of the
states within the levels describing the quantum motion within the Coulomb potential of an electron without spin,
which is $N^2$-fold with $N$ standing for the principal
quantum number of the Coulomb potential problem. In the first case, the degeneracy is due to the rotational invariance
of the three-dimensional  position space, which requires conservation of angular momentum, and demands the total
wave functions of the central potentials
to be simultaneously eigenfunctions of ${\mathbf  L}^2$, and $L_z$,  with ${\mathbf L}$ standing for the angular momentum pseudo-vector, and $L_z$ 
for its $z$-component.   As long as ${\mathbf L}^2$ acts as the Casimir
invariant of the $so(3)$ algebra, the degeneracy with respect to the magnetic quantum number (the $L_z$ eigenvalues $m\in \left[-\ell, +\ell\right]$) 
is attributed to the
rotational invariance of the Hamiltonian. The second case is bit more involved in so far as
in order to explain the larger $N^2$-fold degeneracy, one needs to invoke the higher
$so(4)$ symmetry algebra underlying the Coulomb potential problem by accounting for the constancy of the Runge-Lenz vector next to that
of angular momentum \cite{Elliot_Dawber}.

\noindent
The list can be continued be some more examples, a  popular one being
the case of the P\"oschl-Teller potential,  $V_{\mbox{\footnotesize PT}}=-\frac{\lambda (\lambda +1)}{\cosh^2\eta} $.
In \cite{WuAlhassid} it has been noticed that the free quantum motion 
on the  one-sheeted two-dimensional hyperboloid, $x^2+y^2-z^2=1^2$, an $AdS_2$ space \cite{Moschella},
transforms, upon an appropriate change of variables, into the  one-dimensional (1D) Schr\"odinger equation
with same  potential. As long as the kinetic-energy Laplace-Beltrami operator on the hyperboloid  is
proportional to the Casimir operator of the  $so(2,1)$ isometry algebra of the $AdS_2$ surface, the Hamiltonian of 
the free motion on the curved space
is equivalent (up to additive constant) to the aforementioned Casimir invariant and
the spectrum of the potential under discussion correspondingly exhibits spectral patterns characteristic for $so(2,1)$.
In consequence, also the spectrum of the related 1D-Schr\"odinger equation with the P\"oschl-Teller  potential will be
classified according to the irreducible representations of the same algebra and will  carry patterns identical to those of  the free
motion on $AdS_2$.

\noindent
The idea, that the spectrum of a Schr\"odinger equation with a given potential exhibits certain Lie algebraic degeneracies
because in some appropriately chosen variables it becomes identical to the eigenvalue problem of                 
the Casimir operator  of the isometry algebra of a curved surface, 
the former not necessarily being  unitarily equivalent to its canonical representation,  
has been further elaborated and generalized by many authors  (see \cite{AsimG_2} for a review). It  has become known
in the literature  under  the name of ``symmetry algebra of a potential'', or, simply, ``potential algebra''.
The potential algebra concept  attributes to an underlying
algebraic symmetry the degeneracy patterns of a potential. Predominantly, the $su(1,1)$ symmetry of the Natanzon-class potentials
has been extensively 
studied within this context,
for example in refs.~\cite{Barut},  \cite{Levai},  \cite{mota}, \cite{Nemy}, among others, but also the $so(4)$ symmetry of the
trigonometric P\"oschl-Teller potential \cite{Quesne}, on the one side, and of the trigonometric Rosen-Morse potential
\cite{Adrian}, \cite{MolPhys} on the other side,  has been paid due attention.
In effect,  the observation of degeneracies  in the spectrum of a given potential problem that appear patterned after a known 
regular Lie algebraic symmetry,  as a rule awakes the expectation that the very same algebra may 
determine the symmetry  of the interaction in question. 

\noindent
In the literature on exactly solvable potentials,
degeneracy is usually interpreted as a consequence of some underlying potential algebra. 
Yet,  it is well known a fact that at the same time
quantum mechanics successfully describes  also  phenomena  of degeneracy without symmetry, as is the violation of the 
non-crossing rule in the correlated electron system of Benzene described by means of the Hubbard Hamiltonian \cite{Heilmann}, or the 
detection of resonance degeneracies in a double well potential \cite{Mondragon}. Such type of degeneracies  are ordinarily termed to as 
accidental, better, fortuitous. An indispensable text on the aspects of degeneracy without symmetry  is provided by \cite{Berry} 
within the context of quantum chaotic motion. However, one should keep in mind that  
conditioning degeneracy by symmetry is not exclusive to regular Lie symmetries alone. Especially in non-linear systems,
degeneracy can find explanation in terms of quantum group symmetries, obtained by deforming ordinary Lie algebras,
as observed for degenerate Landau levels 
describing plane motion within a constant magnetic field of a charged massive spin-less particle  \cite{Jellal}. 
Recapitulating the literature on exactly solvable potentials,  
degeneracies  so far have been either associated with eigenvalue problems (in properly chosen variables)
of differential Casimir operators of Lie symmetry algebras, regular or deformed,  or, with the absence of such.

We here draw attention to a different option in showing that degeneracy conservation by perturbation
can also be understood as equivalence between the matrix elements of an interaction Hamiltonian
and the matrix elements of finite polynomials  of a Casimir operator of a properly identified Lie algebra,
the polynomial coefficients  being  interaction (potential)  specific.
Such is the case of the perturbation  of the free quantum motion on $S^3$ by the two-parameter potential,
\begin{equation}
V_{S^3}(\chi)=b^2{\sec ^2\chi} -b(2 \ell+1)\tan\chi \sec\chi.
\label{Gleichung1}
\end{equation}
Here, $\ell$ is discrete natural number, while the external scale introduced by the  parameter
$b$ is continuous.

The paper is structured as follows. In the next section we briefly review for the sake of self-sufficiency of the presentation,
the 1D Schr\"odinger equation with Scarf I with the emphasize on its hydrogen-like degeneracies for $\ell$ non-negative integer,
and review the proof of an $so(4)$ potential algebra for the case of two quantized parameters.  
In section 3 we place the Scarf potential problem on $S_{\mbox{\footnotesize R}}^{d+1}$ for any $d$  although then we 
focus without loss of generality on $S_{\mbox{\footnotesize R}}^3$ for the sake of concreteness.  We
 show that  the wave functions, $\psi_{ N \ell m}(\chi, \theta,\varphi)$, of Scarf I
(with $N$ standing for the principal quantum number, $N=(\ell +n +1)$, and 
$n$ denoting the nodes of the wave function)
do not transform irreducibly under  $so(4)$ 
because they turn to be  mixtures of representation functions transforming according to $(K +1)^2$-dimensional 
carrier spaces of a properly identified $so(4)$ algebra  realization,  
with the value of the 4d angular momentum, $K$, varying as, $ K \in \lbrack {\ell=(N-1-n)}, N-1 \rbrack\in {\mathbf N}$.
The above decompositions are simultaneously expansions in
 power series in the perturbation parameter, $b$, in which carrier spaces of dimensionality $(K+1)^2$ contribute up to
the order ${\mathcal O}\big(b^{N-1-K}\big)$. Also there we present the generalization of the notion of a Hamiltonian from a 
{\it single differential  Casimir operator\/} (potential algebra concept \cite{WuAlhassid})
to an equivalent diagonal matrix form whose elements are  potential specific  {\it finite polynomials\/} 
of the aforementioned Casimir operator (``dynamics governed by generalized  symmetry algebra invariants'', 
abbreviated, ``dynamical symmetry''  \cite{Iach}). 
We  employ the latter concept in the  explanation of  the hydrogen-like degeneracy patterns 
in the spectrum of the Scarf I Hamiltonian.   
The closing section contains the summary  of the results and discusses the perspectives.

\section{The trigonometric Scarf potential and its degeneracy patterns }
The (periodic) trigonometric Scarf potential is of frequent use in the description of di-atomic--,  and poly-atomic molecules in
solid-state physics, on the one side,  or in di-molecular and poly-molecular systems in physical chemistry, on the other. 
In several quantum systems it simulates reasonably well the average effect  exercised by the inter-atomic(intermolecular) interactions 
on a single atom (molecule). The potential is characterized by two parameters only, is exactly solvable, and easy to handle with by 
computational soft-wares, all advantages that make it interesting to both theoretical studies and applications.
Specifically in the present study, we focus on the  peculiarity  that the spectrum of Scarf I in (\ref{Gleichung1}) exclusively depends on 
the $\ell $ parameter alone,  while the importance of the $b$ parameter confines to the level of the  wave functions. 
{}For non-negative integer $\ell$-values, the latter spectrum  
shows typical hydrogen-like degeneracy and one expects the potential
to have the geometric $so(4)$ as a potential algebra. In such a  case the Scarf I Hamiltonian would be linear in the Casimir operator of the
$so(4)$ algebra in a properly designed representation. 
This is true only restrictively,  namely, only if the second parameter, $b$, which is irrelevant to the spectrum,
has been  properly quantized too. This case has been studied in the literature in great detail and the understanding has been 
gained  that the corresponding wave functions transform as genuine $so(4)$ representation functions \cite{Levai}. 
However, for continuous $b$ values this is not to be so, though the spectrum, 
in remaining unaffected, still keeps exhibiting those very  same $so(4)$ degeneracy patterns.
Therefore, the case of Scarf I with one quantized and one continuous parameter may provide an intriguing 
template for studying the phenomenon of possibly observing  Lie-algebraic degeneracies 
without an underlying geometric algebraic symmetry of the potential. 
To illuminate this issue, is the goal of the present study.

\subsection{The general 1D Schr\"odinger equation with the two-parameter  Scarf I  potential}
The 1D Schr\"odinger Hamiltonian, $H_{\textup{ScI}}(\chi)$, with the trigonometric Scarf potential, here denoted by
$V_{\textup{ScI}}(\chi)$,  
and  its exact solutions \cite{Levai} are very well known and given 
(here in dimensionless units $\hbar^2/2MR^2=1$) by 
\begin{eqnarray}
H_{\textup{ScI}}(\chi)\,U\left( \chi \right)
&=&\left[- \frac{{\mathrm d}^2}{{\mathrm d}\chi^2}
+ V_{\textup{ScI}}\left(\chi \right)\right]\,
U\left( \chi \right)=
\epsilon \, U \left( \chi \right),\label{Schr1}\\
 V_{\textup{ScI}}\left(\chi \right)&=&
\frac{b^2+ a(a+1)}{\cos^2 \chi } -\frac{b(2a+1)\tan \chi }{\cos \chi },
\label{Schr_prmts}\\
 b^2=\frac{2MR^2B^2}{\hbar^2},&\quad& a(a+1)=\frac{2MR^2}{\hbar^2}A(A-1),
\label{Schr2}\\
U \left( \chi \right)&=&{\mathbf F}^{-1}(\chi)\,
\cos^{a+1}\chi \, P_n^{a-b+\frac{1}{2}, a+b+\frac{1}{2}}\left(\sin \chi  \right),
\label{Schr3}\\
{\mathbf F}^{-1}(\chi)&=&\left(\frac{1+\sin \chi  }{1-\sin \chi }\right)^{\frac{b}{2}}=e^{-b\tanh^{-1}\sin\chi},
\label{Schr4}\\
 \quad \epsilon = (a+n+1)^2,  &\quad& \epsilon=\frac{ 2ME R^2}{\hbar^2}.
\label{Scarfs_pot}
\end{eqnarray}
The Scarf I potential is determined by two-parameters, denoted by $a$ and $b$ when adimensional, 
or, by $A$ and $B$, when carrying the dimensionality of energies.
The dimensionless angular variable $\chi $ is represented 
as, $ \chi =\frac{r}{R}$, where  $r$ is  a distance, $R$ is a suited  matching  length parameter, 
$E$ is the bound state energy in MeV, $\epsilon$ stands for adimensional energy, 
and $P_n^{\alpha,\beta}(\sin\chi)$ are the Jacobi polynomials.

\subsection{The $so(4)$ algebra of Scarf I with two quantized parameters}
The expression for the energy, $\epsilon$, in (\ref{Scarfs_pot}) is such that for a non-negative integer $a=\ell\in {\mathbf N}$,
the spectrum exhibits a hydrogen-like  $N^2=(\ell+n +1)^2$-fold degeneracy which is characteristic for
an algebraic $so(4)$ symmetry.  This observation suggests 
$H_{\textup{ScI}}(\chi)$ to behave as a differential Casimir operator of a properly constructed  $so(4)$ algebra. 
To general proof of a symmetry of an interaction requires to 
\begin{itemize}
\item identify the symmetry algebra of the emerging degeneracy patterns (recognized as $so(4)$ in the case of interest),
\item confirm irreducibility of the wave functions under transformations of same algebra.
\end{itemize}
{}For the potential under discussion this  program has been executed in  \cite{Levai}.
The line of reasoning is based on the fact that for discrete (quantized)  parameters, 
$a=m-\frac{1}{2}$, and $b=m^\prime $,  the Jacobi polynomials in (\ref{Schr3}) defining the Scarf wave functions become,
\begin{eqnarray}
 P_n^{a-b+\frac{1}{2}, a+b+\frac{1}{2}}\left(\sin \chi  \right)\longrightarrow
 P_n^{m-m^\prime, m+m^\prime}\left(\sin \chi  \right).
\label{Wigner_pre}
\end{eqnarray}
Then, the wave function $U(\chi)$ in (\ref{Schr3}) allows for a factorization of  Wigner's $d^{j=n-m}_{mm^\prime}$ functions, 
the representation functions of an  $su(2)$ algebra, according to,
\begin{eqnarray}
U \left( \chi \right)&=&{\mathbf G}^{-1}(\chi)\, d^{j=n-m}_{m^\prime m}(\sin\chi),\nonumber\\
{\mathbf G}^{-1}(\chi)&=&{\mathcal N}{\mathbf F}^{-1}(\chi) \cos^{\frac{1}{2}}\chi
\tan ^{m^\prime} \frac{\chi}{2},\quad \chi\longrightarrow \frac{\pi}{2}-\chi,
\label{Chri_G}
\end{eqnarray}
with ${\mathbf F}^{-1}(\chi)$ from (\ref{Schr4}), and ${\mathcal N}$ being  a normalization constant.
Such wave functions behave as representation functions of the following similarity transformed  canonical rotational algebra, 
\begin{equation}
{\widetilde {\mathbf J}}^2(\chi,\varphi)={\mathbf G}^{-1}(\chi){\mathbf J}^2(\chi,\varphi){\mathbf G}^{-1}(\chi),
\label{L2tild}
\end{equation}
with $\chi$ now playing the r\'ole of the ordinary polar angle.
This algebra has not been worked out explicitly in \cite{Levai} but its ladder operators have been
properly identified by inspection on the basis 
of the Schr\"odinger ladder operators,  $A^\pm=-\partial_\chi \pm W_{ScI}$, with
$W_{ScI}=-(a -1) \tan\chi +b\sec\chi$ standing for the super-potential of $V_{ScI}$, and
exploiting their property to factorize   $H_{ScI}$.
In a similar way, a second  $su(2)$ algebra has been constructed on the basis of the 
super-symmetric partner, ${\widetilde H}_{ScI}(\chi)$.
The discrete parameters, $m$, and $m^\prime$  have been introduced  
as auxiliary phases into the wave functions according to, $U(\chi)\to \exp(im\alpha +im^\prime \beta) U(\chi)$. 
In effect,  two sets of ladder operators have been designed, in turn labeled as left (L) and right (R) handed.
Their algebras  have  then been closed to $su(2)_L$, and $su(2)_R$
by $(-i\partial_{\alpha})$,  and  $(-i\partial_\beta)$, respectively.   
As a result, left-handed $su_L(2)$,  and right-handed $su_R(2)$ algebras have been designed and their direct sum,
 $su_L(2)\oplus su_R(2)$, identified with the algebra of the universal cover
 $SU_L(2)\otimes SU_R(2)$ of the group $SO(4)$. This algebra locally is isomorphic to $so(4)$, 
i.e.  $su_L(2)\oplus su_R(2)\simeq so(4)$.
Then the  Scarf I Hamiltonian has been cast in the form of
the corresponding Casimir operator, denoted by ${\mathcal C}_2^{J_LJ_R}$,
that expresses in terms  of the respective squared left (L) and right (R) handed
angular momentum operators, ${\mathbf J}_L^2$, and ${\mathbf J}_R^2$ as
 \begin{eqnarray}
{\mathcal C}_2^{J_LJ_R}=2({\mathbf J}^2_L +{\mathbf J}_R^2)=2{\mathcal K}^2,
\label{C2}
\end{eqnarray}
with ${\mathcal K}^2$ standing for the operator of the squared 4D angular momentum.
The above scheme has been independently employed to establish the $so(4)$ symmetry of the trigonometric 
P\"oschl-Teller potential by Quesne in \cite{Quesne}. We here closely followed precisely this very reference. 
Using the algebra locally isomorphic to $so(4)$ from above has the advantage that one generates both 
the common and the projective representations. As a reminder, the group $SO(4)$ is the quotient of 
the universal covering $SU(2)_L\otimes SU_R(2)$ by
the  center $Z$, i.e. $SO(4)\simeq SU_L(2)\otimes SU_R(2)/Z$.
It is interesting to notice that $SO(4)$  can alternatively be viewed as the quotient of the Euclidean group $E^+(4)$ by the group of translations, i.e.
$SO(4)\simeq E^+(4)/T$, which would provide a different technique for the treatment the above problem \cite{Euclid}.

To recapitulate, the requirement on algebraic $so(4)$ symmetry of the  Scarf I Hamiltonian imposes on its parameters
stringent conditions of quantization.
 The property of the Schr\"odinger ladder operators to partake a closed Lie algebra under certain restrictions on the values of
the potential parameters,  provides any time that such is possible, a powerful method for  
the algebraic description of  various quantum mechanical  problems  such as those related to  the 
Coulomb--, the Harmonic-Oscillator and other interactions \cite{mota}.  

However, without the quantization of the $b$ parameter, the equality in (\ref{Wigner_pre}) is no longer valid
and one is left with 
\begin{equation}
U \left( \chi \right)={\mathbf F}^{-1}(\chi)\,
\cos^{\ell+1}\chi \, P_n^{\ell-b+\frac{1}{2}, \ell+b+\frac{1}{2}}\left(\sin \chi  \right),
\label{nash}
\end{equation}
where from now onwards  $a$ will be quantized to non-negative integer as $a=\ell$.
The Jacobi polynomials in the latter equation are such that neither Wigner functions, nor
Gegenbauer polynomials can be in general factorized, a circumstance that strongly points towards serious
difficulties in the construction of an explicit similarity transformation of the canonical geometric
$so(4)$ algebra towards the Scarf I Hamiltonian with one quantized and one continuous parameter.   

In effect, the status of $so(4)$ as a symmetry algebra of the Scarf I Hamiltonian is  no longer
obvious, although  the spectrum, in being independent of $b$, remains unaltered. 
It is one of the goals of the present study to examine the relationship between the Scarf I Hamiltonian and
\begin{equation}
{\widetilde {\mathcal K}}^{2}= {\mathbf F}^{-1}{\mathcal K}^{2}{\mathbf  F},
\label{bo}
\end{equation}
as suggested by the wave function in (\ref{nash}).
{}For that purpose, placing the problem on a hyper-spherical surface
turns to be helpful.

\section{Perturbing the free quantum motion on $S^{d+1}$ by Scarf I with one quantized parameter}

The free motion on the $(d+1)$ dimensional unit hypersphere, to be denoted by $S^{d+1}$ embedded within a $(d+2)$ dimensional Euclidean space, $E_{d+2}$, is given by
\begin{eqnarray}
\Delta_{S^{d+1}} Y_{K_ {d+2}K_{d+1}... {\ell}m}(\chi, \eta,.., \theta, \varphi)= K_{d+2}\left( K_{d+2}+ d \right)
Y_{K_ {d+2}K_{d+1}... {\ell}m}(\chi, \eta...\theta,\varphi),
\label{HSy_1}
\end{eqnarray}
where $\Delta_{S^{d+1}}$ is the Lapalce-Beltrami operator on the surface under consideration, defined as,
\begin{eqnarray}
\Delta_{S^{d+1}}=-\frac{1}{\cos^d\chi}\frac{\partial }{\partial \chi }\cos^d \chi 
\frac{\partial}{\partial \chi } +\frac{{\mathbf K}_{d+1}^2}{\cos^2\chi}.
\label{LpBltr}
\end{eqnarray}
Here, ${{\mathbf K}_{d+1}^2}$ stands for the squared angular momentum operator in the Euclidean space, $E_{d+1}$, of one less dimension,
$\chi$, $\eta$,...,$\theta\in \left[ -\frac{\pi}{2},+\frac{\pi}{2}\right]$ are polar angles, $\varphi \in \left[0,2\pi \right]$ 
is the standard azimuthal angle,
  $Y_{K_ {d+2}K_{d+1}... {\ell}m}(\chi,\eta,...,\theta, \varphi)$ 
are the hyper-spherical harmonics on $S^{d+1}$, and $K_{d+2-t}$ stand for the angular momentum values within an Euclidean 
spaces of $t$ less dimensions, i.e in $E_{d+2-t}$. Finally,  $\ell $, $m$ are in turn the standard $E_3$ and $E_2$  angular momenta.  
Confining to the quasi-radial motion with wave function, denoted by $R_{K_{d+2}K_{d+1}}(\chi)$,
and changing variable as,
\begin{eqnarray}
R_{K_{d+2}K_{d+1}}(\chi )=\frac {U_{K_{d+2}K_{d+1}}(\chi)}{\cos^{\frac{d}{2}}\chi }, 
\label{var_change}
\end{eqnarray}
amounts to the following one-dimensional Schr\"odinger equation,

\begin{eqnarray}
\left[ -\frac{
{\mathrm d} ^2}
{
{\mathrm d}
\chi^2} +
\frac{\left(K_{d+1}+\frac{d-1}{2}\right)^2-\frac{1}{4}}{\cos^2\chi}\right]U_{\footnotesize{K_{d+2}K_{d+1}}}(\chi) &=&
\left[ K_{d+2}\left(K_{d+2}+d\right) +\frac{d^2}{4}\right]U_{\footnotesize{K_{d+2}K_{d+1}}}(\chi).\nonumber\\
\label{1d_Hyps}
\end{eqnarray}
Comparison  to (\ref{Schr_prmts}) reveals  (\ref{1d_Hyps})
as the Scarf I potential problem for $b=0$,
and
\begin{equation}
a=K_{d+1}+\frac{d-1}{2} -\frac{1}{2},
\label{a_genrl}
 \end{equation}
with $a$ either integer, or semi-integer.
In this manner, the explicit  $so(d+2)$ potential algebras of the $\sec^2$  interaction of the one-dimensional Schr\"odinger equation
have been made manifest. In now switching to the full Scarf I potential in (\ref{Schr1})--(\ref{Scarfs_pot})
amounts to the following  perturbed  motion on $S^{d+1}$,

\begin{eqnarray}
\left[
-\frac{1}{\cos^d\chi}\frac{\partial }{\partial \chi }\cos^d \chi 
\frac{\partial}{\partial \chi } +\frac{b^2 + K_{d+1}^2 }{\cos^2\chi}
-\frac{b(2 K_{d+1}+ d-1)\tan \chi  }{\cos\chi }
\right]\phi_{K_{d+2}K_{d+1}} (\chi)&&\nonumber\\
=\epsilon_{K_{d+2}}\phi_{K_{d+2}K_{d+1}}(\chi)&&\nonumber\\
\epsilon_{K_{d+2}}=K_{d+2}\left(K_{d+2}+d\right)+ \frac{d^2}{4}.&&
\label{Pert_Sph}
\end{eqnarray}

Correspondingly, the solutions to (\ref{Pert_Sph})  are read off from (\ref{Schr3}) as

\begin{eqnarray}
\phi_{K_{d+2}K_{d+1}}(\chi)= \frac{U(\chi)}{\cos^{\frac{d}{2}}\chi }, \quad U(\chi)=
{\mathbf F}^{-1}(\chi)\cos ^{K_{d+1} +\frac{d-1}{2}+\frac{1}{2}  }\chi P_n^{K_{d+1}+ \frac{d-1}{2}-b, K_{d+1}+ \frac{d-1}{2}+b }(\sin\chi), 
\label{wafu_Sd}
\end{eqnarray}
with $a$ from (\ref{a_genrl}).
Comparison to (\ref{HSy_1}) shows that the perturbation retains the degeneracy  patterns of the free motion
in {\it any dimension} and rises the question on the symmetry of the full trigonometric Scarf potential.

In the following, we shall focus on $S^3$, setting $d=2$, for concreteness, and without  loss of generality.
In so doing, $K_{4}$ becomes the four-dimensional angular momentum value, to be denoted  by $K$ only,
while $K_{2+1}$ is no more but the ordinary angular momentum, $\ell$.
With that, i.e. for $a=\ell$,  the perturbation potential announced in (\ref{Gleichung1}) in the introduction becomes,
\begin{equation}
V_{S^3}(\chi) =\frac{b^2}{\cos^2\chi } -\frac{b(2 \ell+1)}{\cos\chi}\tan\chi.
\label{S3potential}
\end{equation}
It is that very potential that will be referred to from now onwards  as Scarf I on $S^3$.
Apparently, for $b=0$   the free quantum motion on the curved surface under consideration is recovered.
{}For $d=2$, the equation (\ref{Pert_Sph}) can also be viewed as the 4D quantum mechanical rigid rotator perturbed by $V_{{S^3}}(\chi)$,
a problem of interest to di-atomic-- or di-molecular systems.  
In  furthermore recalling the relationship between the 
Laplace-Beltrami operator and the operator of the squared 4D angular momentum,
${\mathcal K}^2$, a Casimir invariant of the $so(4)$ isometry algebra of $S^3$,
\begin{equation}
-\Delta_{S^3}= {\mathcal K}^2,
\label{Ksquared_free}
\end{equation}
allows to cast the free quantum motion on the unit hyper-sphere $S^3$ in terms of the ${\mathcal K}^2$ eigenvalue problem as 
\begin{equation}
\left( {\mathcal K}^2+1\right)Y_{K\ell m}(\chi, \theta, \varphi)=
 (K+1)^2 \, Y_{K\ell m}(\chi, \theta, \varphi).
\label{geod}
\end{equation}
Here, $Y_{K\ell m}(\chi, \theta, \varphi)$ stand for the well-known 4D hyper-spherical harmonics, 
and with $K\in [0,\infty)$, $\ell \in [0,K]$, and $m\in [-\ell,+\ell]$. The 4D hyper-spherical harmonics 
are the representation functions of the isometry  $so(4)$ algebra,
which describe $(K+1)^2$-dimensional $so(4)$ carrier spaces and are 
defined according to,
\begin{eqnarray}
Y_{K\ell m}(\chi, \theta, \varphi)={ S}_{K\ell}(\chi) \, Y_{\ell}^m(\theta,\varphi), 
\quad S_{K\ell}(\chi)=\cos^{\ell}\chi \, {\mathcal G}_{n=K-\ell}^{\ell+1}(\sin\chi).
\label{quasirad}
\end{eqnarray}
Here, ${\mathcal G}_{n=K-\ell}^{\ell+1}(\sin\chi) $ stand for the Gegenbauer polynomials.
The $S_{K\ell}(\chi)$ functions are sometimes referred to as the ``quasi-radial'' functions of the free motion \cite{Mardoyan}.

Here, ${\mathcal K}^2$ is expressed in terms of the six generators  $J_i$ and $A_i$ with $i=1,2,3$, 
spanning the $so(4)$ algebra \cite{Englefield},
\begin{eqnarray}
\left[ J_i,J_j\right] =i\epsilon_{ijk}J_k,\quad \left[ A_i, A_j \right]=i\epsilon_{ijk}J_k,\quad
\left[ J_i, A_k\right] =i\epsilon_{ijk}A_k,
\label{isomtry_alg}
\end{eqnarray}
as
\begin{eqnarray}
 {\mathcal K}^2&=& 2\,\left( {\mathbf J}^2 +{\mathbf A}^2\right). 
\end{eqnarray}

In terms of (\ref{Ksquared_free}),
the equation of the perturbed motion on $S^3$  which we will be dealing with here, takes the following final shape,
\begin{eqnarray}
{\mathcal H}_{\mbox{\footnotesize Sc}}(\chi)\psi_{N \ell m}(\chi, \theta,\varphi) &=&\epsilon_N \, \psi_{N \ell m}(\chi, \theta, \varphi), \nonumber\\
{\mathcal H}_{\mbox{\footnotesize Sc}}(\chi)&=&{\mathcal K}^2 +1 + V_{S^3}(\chi),\nonumber\\
{\mathcal K}^2& =& -\frac{1}{\cos^2\chi}\frac{\partial}{\partial \chi }\cos^2 \chi 
\frac{\partial}{\partial \chi } +\frac{{\mathbf L}^2}{\cos^2\chi},\nonumber\\
\psi_{N \ell m}(\chi, \theta,\varphi )&=&\phi_{N \ell}(\chi) \, Y_{\ell}^m(\theta,\varphi)\nonumber\\
&\equiv& {\mathbf F}^{-1}(\chi ) \,
\cos^{{ \ell}}\chi \, P_{n={ N-1}-{ \ell}}^{\alpha,\beta} (\sin\chi) \, Y_{\ell}^m( \theta, \varphi),\nonumber\\
N=n+\ell+1,\quad \alpha=\ell-b+\frac{1}{2}, &\quad&\beta=\ell+b+\frac{1}{2},
\label{ScarfIS3}
\end{eqnarray}
with $V_{S^3}$ from (\ref{S3potential}), and ${\mathbf F}^{-1}(\chi)$ from (\ref{Schr4}).
The  energy excitations are, 
\begin{equation}
\epsilon_{ N }=N^2, \quad N\in {\mathbf N}, \quad N\in \lbrack 1,\infty ).  
\end{equation} 

The remarkable aspect of the perturbation by the trigonometric Scarf potential is that despite the drastic change in 
the quasi-radial wave functions from unperturbed, $S_{K\ell}(\chi)$ in (\ref{quasirad}), to perturbed,  
$ \phi_{N\ell}(\chi)$ in (\ref{ScarfIS3}) according to,
\begin{eqnarray}
S_{K\ell}(\chi)=\cos^{\ell}\chi \, {\mathcal G}_{n=K-\ell}^{\ell+1}(\sin\chi)
&\longrightarrow& \phi_{N \ell}(\chi)= e^{-b\tanh^{-1}\sin\chi} \,
\cos^{{ \ell}}\chi \, P_{n={ N-1}-{ \ell}}^{\alpha,\beta} (\sin\chi) ,
\label{change}
\end{eqnarray}
 its spectrum remains independent of the external-scale introducing parameter $b$, which can be as well infinitesimally small, as  finite.
This spectrum is still characterized by that very same  $N^2$-fold degeneracy of the states in a level, just as the hydrogen atom, 
and formally copies the $({\mathcal K}^2+1)$-eigenvalue problem.
{}From the equation (\ref{change}) one immediately reads off that
for the particular case of the parameter $\ell$ taking its maximal value of $\ell=(N-1)$ (it includes the ground state, $(N-1) =\ell=0$),
the polynomials on both sides are of zero degree, i.e. constants,  and one encounters
equality between the  Scarf I solutions on $S^3$, on the one side, and 
the ${\mathbf F}^{-1}(\chi)$ transformed  representation functions of the isometry $so(4)$ algebra, on the other, namely, 
\begin{equation}
\psi_{N, \ell=(N-1), m}(\chi, \theta, \varphi)={\mathbf F}^{-1}(\chi)\,Y_{K=(N-1), \ell=(N-1), m}(\chi ,\theta, \varphi)\equiv
 {\widetilde Y}_{K=(N-1),  \ell=(N-1), m}(\chi,\theta,\varphi),
\label{gst}
\end{equation} 
holds valid. Notice that ${\widetilde Y}_{K=(N-1),  \ell=(N-1), m}(\chi,\theta,\varphi)$ behave as  
representation functions of an $so(4)$ algebra similarity transformed to,
${\mathbf F}^{-1}{\mathcal K}^2{\mathbf F}={\widetilde {\mathcal K}^2}$.
{}For this particular case the solutions of the perturbed quantum motion on $S^3$ under discussion result $so(4)$ symmetric, 
though the algebra is in a representation that is unitarily nonequivalent to the hyper-spherical one.
The equation (\ref{gst}) is suggestive of a relationship  between the Scarf I Hamiltonian and
a similarity transformation of the geometric $so(4)$ algebra in (\ref{isomtry_alg}) 
by the exponential function ${\mathbf F}^{-1}(\chi)$ in (\ref{Schr4}), which also formed part of the design of the
explicit potential algebra of Scarf I with two quantized parameters in (\ref{Chri_G}).
The next subsection is devoted to the construction of the transformed algebra (\ref{bo})
(with ${\mathcal C}_2^{J_L J_R}$ being replaced by the squared 4D angular momentum operator, ${\mathcal K}^2$) 
and to the comparison of its Casimir operator to the Scarf I Hamiltonian in (\ref{ScarfIS3}).

\subsection{Geometric so(4) symmetry loss in  the Scarf I potential problem  on $S^3$ without 
lifting the degeneracy of the free quantum motion}

We are interested in calculating the class of representation functions of the $so(4)$ algebra 
to which\newline  ${\widetilde Y}_{K=(N-1),  \ell=(N-1), m}(\chi,\theta,\varphi)$ in (\ref{gst}) belong.
{}For this purpose we consider the algebra spanned by the set of elements,
\begin{eqnarray}
\widetilde{ J}_i ={\mathbf F}^{-1}J_i{\mathbf F}, &\quad &
\widetilde {A}_i={\mathbf F}^{-1}A_i{\mathbf F},\quad i=1,2,3,\nonumber\\
{\widetilde {\mathcal K}}^2={\mathbf F}^{-1}{{\mathcal K}^2}{\mathbf F}, &\quad & {\widetilde {\mathcal K}}^2= 2\,
\sum_{i=1}^{i=3}\left( {\widetilde J}^2_i +{\widetilde A}^2_i\right),
\label{Cas_new_rep}
\end{eqnarray}
with $ {\mathbf F}^{-1}(\chi)$ from (\ref{Schr4}).
The corresponding representation functions will be termed to as exponentially rescaled hyper-spherical harmonics, defined as,
\begin{eqnarray}
{\widetilde Y}_{K\ell m}(\chi,\theta,\varphi)&=& {\mathbf F}^{-1}(\chi)\, S_{K\ell}(\chi)\, Y_\ell^m(\theta,\varphi)
=\widetilde{S}_{K\ell}(\chi)\,Y_\ell^m(\theta,\varphi),\nonumber\\
\widetilde{S}_{K\ell}(\chi)&=& {\mathbf F}^{-1}(\chi)\, S_{K\ell}(\chi).
\label{new_repr_fu}
\end{eqnarray}
Next we calculate the similarity transformed Casimir operator,
$\left[ {\widetilde {\mathcal K}}^2+1\right]=\left[ {\mathbf F}^{-1}(\chi)
{{\mathcal K}^2}{\mathbf F}(\chi)+1\right]$,
together with  its action on the  ${\widetilde S}_{K\ell}(\chi)$, functions i.e.

\begin{equation}
\left[ {\widetilde {\mathcal K}}^2+1\right]{\widetilde S}_{ K\ell}(\chi) =
\left[ {\mathbf F}^{-1}(\chi){{\mathcal K}^2}{\mathbf F}(\chi)+1\right]             
{\widetilde S}_{ K \ell}(\chi).
\label{Anlauf}
\end{equation}
In so doing we find the following general expression (see \cite{WuAlhassid} for a similar procedure),
\begin{eqnarray}
\left[ 
{\mathbf F}^{-1}(\chi) {\mathcal K}^2{\mathbf F}(\chi)+1\right]{\widetilde S}_{ K \ell}(\chi)&=&
\left[{{\mathcal K}}^2+1 +{\mathcal V}(\chi)\right]{\widetilde S}_{ K\ell}(\chi) ,\nonumber\\
{\mathcal V}(\chi)&=&
-{\mathbf F}^{-1}(\chi) 
\left[ \frac{\partial ^2{\mathbf F}(\chi)}{\partial \chi^2}
- 2\tan \chi \frac{\partial {\mathbf F}(\chi)}{\partial \chi}\right] {\widetilde S}_{ K \ell}(\chi)\nonumber\\
&-&2{\mathbf F}^{-1}(\chi)\frac{\partial {\mathbf F}(\chi)}{\partial\chi}\frac{\partial {\widetilde S}_{ K \ell}(\chi)}{\partial \chi}.
\label{trans_formed}
\end{eqnarray}
The latter equation makes manifest that similarity transformations of algebra invariants generate 
potentials ${\mathcal V}(\chi)$ which  in general contain gradiental terms.

{}For the specific form of  the function ${\mathbf F}(\chi)$ defined  in (\ref{Schr4}) one finds,
\begin{eqnarray}
-{\mathbf F}^{-1}(\chi) \frac{\partial ^2{\mathbf F}(\chi)}{\partial \chi^2}
&=&\left( -\frac{b^2}{\cos^2\chi} +b\frac{\tan\chi}{\cos \chi }\right),
\label{step12}\\
2\tan \chi{\mathbf F}^{-1}(\chi) 
 \frac{\partial {\mathbf F}(\chi)}{\partial \chi}
&=&-2b\frac{\tan\chi}{\cos\chi},
\label{step11}\\
-2{\mathbf F}^{-1}(\chi)\frac{\partial {\widetilde S}_{ K\ell}(\chi)}{\partial \chi}
\frac{\partial {\mathbf F}(\chi)}{\partial\chi}
&=&\frac{2b^2}{\cos^2\chi}{ \widetilde S}_{K\ell}(\chi)  -\frac{2b\ell}{\cos\chi}\tan\chi {\widetilde S}_{K\ell}(\chi)\nonumber\\
&+ &\frac{2b}{\cos\chi}
{\mathbf F}^{-1}(\chi) \cos^\ell\chi \frac{\partial {\mathcal G}_n^{\ell +1}(\sin\chi)}{\partial \chi},
\label{step1}
\end{eqnarray}
where use has been made of
\begin{equation}
\frac{\partial {\mathbf F}(\chi )}  {\partial \chi }=-
\frac{b}{\cos\chi}{\mathbf F}(\chi).
\end{equation}
Putting all together and
back-substituting the  eqs.~(\ref{step12})--(\ref{step1}) into (\ref{trans_formed}) yields
\begin{eqnarray}
\left({\widetilde {\mathcal K}}^2+1 \right){\widetilde S}_{K\ell}(\chi)&=&
\left[ {\mathcal K}^2+1+
\frac{b^2}{\cos^2\chi } -\frac{b(2 \ell+1)}{\cos\chi}\tan\chi\right] {\widetilde S}_{K\ell}(\chi)\nonumber\\
&&+
\frac{2b}{\cos\chi}{\mathbf F}^{-1}(\chi)\cos^\ell\chi \frac{ \partial {\mathcal G}_n^{\ell+1}(\sin \chi)}{\partial \chi},
\label{pot_prod}
\end{eqnarray}
meaning that the similarity transformation of ${\mathcal K}^2$ by ${\mathbf F}^{-1}(\chi)$
gives only partially rise to the anticipated  trigonometric Scarf potential, $V_{S^3}(\chi)$ in (\ref{S3potential}), 
the rest being a gradiental term, 
\begin{eqnarray}
\left( {\widetilde {\mathcal K}}^2+1\right){\widetilde S}_{K \ell}(\chi) =
{\mathcal H}_{\mbox{\footnotesize Sc}}(\chi){\widetilde S}_{ K\ell}(\chi)+ 
\frac{2b}{\cos\chi}{\mathbf F}^{-1}(\chi)\cos^\ell\chi 
\frac{\partial {\mathcal G}_n^{\ell +1}(\sin \chi)}{\partial \chi}.
\label{sim_trans1}
\end{eqnarray}
The latter equation 
equivalently rewrites to,
\begin{eqnarray}
(K+1)^2{\widetilde S}_{K\ell}(\chi)
=
{\mathcal H}_{\mbox{\footnotesize Sc}}(\chi){\widetilde S}_{K\ell}(\chi)+\frac{2b}{\cos\chi}{\mathbf F}^{-1}(\chi)\cos^\ell\chi
\frac{ \partial {\mathcal G}_n^{\ell +1}(\sin \chi)}{\partial \chi}.
\label{sim_trans}
\end{eqnarray}
The conclusion is, that   ${\widetilde S}_{Kl}(\chi)$ do not solve the ${\mathcal H}_{\mbox{\footnotesize Sc}}(\chi) $-- eigenvalue problem,
except for the $\ell=(N-1)$ case already discussed in the above equation (\ref{gst}). And vice versa, for the Scarf I solutions,
${\mathcal H}_{\mbox{\footnotesize Sc}}(\chi)\phi_{N\ell}(\chi)=N^2\phi_{N\ell}(\chi)$, one finds,

\begin{eqnarray}
N^2{\phi}_{N\ell}(\chi)
=\left( {\widetilde {\mathcal K}}^2 +1\right)\phi_{N\ell}(\chi) 
-\frac{2b}{\cos\chi}{\mathbf F}^{-1}(\chi)\cos^\ell\chi
\frac{ \partial {P}_n^{\alpha,\beta}(\sin \chi)}{\partial \chi},
\label{sim_trans2}
\end{eqnarray}
that they are no eigenfunctions to $\widetilde{{\mathcal K}}^2$ due to the non-commutativity of the Scarf I Hamiltonian and the
Casimir operator of the transformed algebra,

\begin{equation}
 \left[ {\mathcal H}_{\mbox{\footnotesize Sc}}(\chi), {\widetilde {\mathcal K}}^2\right]\not=0.
\label{non-com}
\end{equation}
\begin{quote}
Our case is that for a general $\ell \not=(N-1) $, the wave functions $\psi_{N \ell m}(\chi, \theta,\varphi)$  in (\ref{ScarfIS3})  
describing  the perturbed motion on $S^3$  behave as mixtures of the type
$\psi_{N\ell m}(\chi,\theta,\varphi)= \Sigma_{K=\ell}^{K=N-1}\,c_{K\ell}(b^{\eta})\,{\widetilde Y}_{K\ell m}(\chi, \theta, \varphi)$, 
with $\eta \in \lbrack 0, n\rbrack$.
Such a property is bound to remain  independent on the parametrization of the hypersphere
by virtue of the model independence of the Lie  algebras.                                                                                                                                                                                                                                                                                                       
\end{quote}

\subsection{The case $\ell=(N -2 )$  as an illustrative example}
We now take a closer look on  ${ \ell }=({ N }-2)$, the state with the next-to highest orbital angular momentum value
within the multiplet,
in which case the wave function of interest is given by
\begin{eqnarray}
\phi_{ N (N -2)}(\chi)=
{\mathbf F}^{-1}(\chi) \cos^{({ N} -2)}\chi P_1^{({ N }-2) -b+\frac{1}{2}, ({ N }-2)+b+\frac{1}{2}}(\sin \chi).
\label{firstwafu}
\end{eqnarray}
The Jacobi polynomial allows for the following decomposition into Gegenbauer polynomials,
\begin{eqnarray} 
P_1^{({ N}-2) -b+\frac{1}{2},\, ({ N }-2)+b+\frac{1}{2}}(\sin\chi)&=& 
-b + \frac{1}{2 }\left(2{N} -1 \right) \sin\chi  \nonumber\\ 
&=& -b\, {\mathcal G}_0^{N-1} (\sin\chi) + \frac{\left(2{N } -1\right)}{4(N-1) }\,{\mathcal G}_1^{N-1} (\sin\chi). 
\label{Jac_Gegenb} 
\end{eqnarray}  

In noticing that by the aid of eq.~(\ref{quasirad}),
\begin{equation} 
\cos^{N -2}\chi \,\, {\mathcal G}_0^{N-1} (\sin\chi) =S_{(N -2)(N -2)}(\chi),\quad 
\cos^{N -2}\chi \,{\mathcal G}_1^{N-1} (\sin\chi) = S_{(N-1) (N-2)}(\chi), 
\end{equation} 
allows to equivalently rewrite eq.~(\ref{firstwafu}) as
\begin{eqnarray}
\phi_{N (N -2)}(\chi)&=& -b\,{\widetilde  S}_{(N-2) (N-2)}(\chi) + 
\frac{(2N -1)}{4 (N-1) }\, {\widetilde S}_{(N-1) (N -2)}(\chi),
\label{BIBONG1}
\end{eqnarray}
with ${\widetilde S}_{K\ell}(\chi)={\mathbf F}^{-1}(\chi)S_{K\ell}(\chi)$ defined in (\ref{new_repr_fu}), and 
standing for the quasi-radial  representation functions of the transformed algebra in
(\ref{Cas_new_rep}).
In consequence, same relationship holds valid at the level of the total Scarf I wave--, and 
${\widetilde {\mathcal K}}^2$ representation  functions,
the exponentially rescaled hyper-spherical harmonics in (\ref{new_repr_fu}),
\begin{eqnarray}
\psi_{N(N -2)m}(\chi,\theta,\varphi)= -b\,{\widetilde Y}_{(N -2)(N -2)m}(\chi, \theta,\varphi )
& +& \frac{(2N- 1)}{4(N-1) }\,{\widetilde Y}_{(N-1) (N-2)m}(\chi,\theta,\varphi).
\label{BIBONG}
\end{eqnarray}

In effect, we observe that 
the lower dimensional $so(4)$ carrier  space, ${\widetilde Y}_{(N -2)(N -2)m}(\chi,\theta,\varphi)$, contributes to 
the order ${\mathcal O}(b^1)$ to $\psi_{N(N-2)m}(\chi,\theta,\varphi)$, i.e.,
\begin{eqnarray}
\psi_{N \ell m}(\chi,\theta,\varphi) &=&\Sigma_{K=\ell}^{K=N-1}c_{K\ell}(b^{N-1 -K}){\widetilde Y} _{K\ell m}(\chi,\theta,\varphi),\quad \ell =N -2,
\label{first_decmp}
\end{eqnarray}
with the expansion coefficients  $c_{K\ell}(b^{N-1 -K})$ being uniquely fixed through the decomposition of the Jacobi into Gegenbauer polynomials.
Therefore,  for  $\ell=(N-2)$, the Scarf I Hamiltonian on $S^3$ deviates to the order ${\mathcal O}(b^1)$ 
from the geometric $so(4)$ algebra Casimir in (\ref{Cas_new_rep}).  

It is straightforward to verify that for any $\ell \not=(N-1) $
the  wave functions of a motion on $S^3$, perturbed by the trigonometric Scarf potential, 
always represent themselves as mixtures of $so(4)$ representation functions corresponding to carrier spaces of 
different dimensionality. Examples are listed in the Table \ref{chi_functions}. 
The generalization of eq. (\ref{BIBONG}) to any ${\ell}$ reads
\begin{equation}
\psi_{ N\ell m}(\chi,\theta,\varphi)=
\sum_{K={ \ell}}^{{ K=N-1 }}c_{K\ell}(b^{\eta}) {\widetilde Y}_{ K \ell m}(\chi,\theta,\varphi), \quad 
\eta \in \left[0, n\right].
\label{decomp}
\end{equation}
Notice summation over the $K$-index defining the dimensionality of the $so(4)$ carrier spaces.
{}For the lowest $\ell $ values,  $\ell =(N-1), (N -2),(N -3)$, one finds $\eta =(N-1 -K)$.
Therefore, the wave functions constituting an $N^2$ degenerate multiplet of Scarf I transform irreducibly solely  under $so(3)$.
Further symmetries are those relevant for any arbitrary degeneracy problem, like $GL(N^2)$, or, $SO(N^2)$, meaning that any 
linear combination of the states within the multiplet is an eigenstate to the Hamiltonian under investigation that 
belongs to same eigenvalue as the basis vectors.

\begin{table}[h]
\begin{center}
        \begin{tabular}{lcp{11.25cm}}
\hline
$\phi_{N {\ell}} (\chi)$&=&$\sum_{K=\ell}^{K=N-1} c_{K\ell}(b^{\eta}){\widetilde S}_{K{\ell}}(\chi) $ \\[3mm]
\hline   \hline
 $\phi_{N (N-1) }(\chi)$ &=& ${\widetilde S}_{(N-1)(N-1)}(\chi)$ \\[3mm] 
 $\phi_{N (N-2)}(\chi)$&=&$\frac{2N-1}{4(N-1)}\,
{\widetilde S}_{(N-1)(N-2)}(\chi)\,-\,b\,
{\widetilde S}_{(N-2)(N -2)}(\chi)$ \\[2.5mm]
$\phi_{N (N-3)}(\chi)$&=&
${1\over 8}\frac{(2N -1)}{(N-2)}\, {\widetilde S}_{(N-1)(N-3)}(\chi)
\,-\,\frac{b}{2}\frac{(N-1)}{(N-2 )}\,{\widetilde S}_{(N-2)(N-3)}(\chi)
\,+\,\frac{b^2}{2}\,{\widetilde S}_{(N-3)(N-3)}(\chi)$ \\[3mm]
 $\phi_{N(N-4)}(\chi)$&=&${1\over 32} \frac{4(N-1)^{2}-1}{(N-3)(N-2)}
{\widetilde S}_{(N-1)(N-4)}(\chi)-
\frac{b}{8}\,\frac{(2N-3) (N-1) }{(N-2)(N -3)}\,{\widetilde S}_{(N-2)(N-4)}(\chi)$ 
\phantom{${\widetilde S}_{(N-2)(N-4)}(\chi)$}
  $\,+\,\frac{b^2}{8}\frac{(2N-3)}{(N -3)}\,{\widetilde S}_{(N-3)(N-4)}(\chi)$ 
  $\,-\,{b \over 24 }\left[\frac{4b^2 (N -2) + (2N -1)}{(N-2)}\right]\,{\widetilde S}_{(N-4)(N-4)}(\chi)$ \\[3mm]
\hline
        \end{tabular}
\end{center}
\caption{Decompositions of  ``quasi-radial'' wave functions, $\phi_{N \ell}(\chi)$,  
of Scarf I in eq.~(\ref{ScarfIS3}) in the basis of the  ``quasi-radial'' parts,
${\widetilde S}_{K\ell}(\chi)$ of the exponentially rescaled hyper-spherical harmonics,  ${\widetilde Y}_{K\ell m}(\chi, \theta, \varphi)$,  
in (\ref{new_repr_fu}). It is well visible that the Scarf I solutions are
mixtures of representation functions describing $so(4)$ carrier spaces of different dimensionality, 
thus making  the geometric $so(4)$ symmetry loss manifest.
The decompositions are simultaneously  finite  power series in the symmetry breaking scale $b$.
Notice that the  leading order $\left(  {\mathcal O}(b^0)\right)$ terms respect the symmetry of the unperturbed motion, as it should be.
In this fashion, a quantitative scheme is elaborated which allows to keep track of the order to which the Scarf I Hamiltonian deviates
from the $so(4)$ Casimir in (\ref{Cas_new_rep}).} 
\label{chi_functions}
\end{table}

Finally, a comment is in order on the reason for which  the perturbation of the quantum motion on $S^3$  
by Scarf I nonetheless happens to conserve the $so(4)$ degeneracies.
Though $\phi_{N \ell}(\chi)$ by themselves do  not behave as $\left({\widetilde{\mathcal K}}^2+1\right)$ eigenfunctions,  the contributions of 
the  gradient term are such that one ends up with the common  $N^2$ eigenvalue and $N^2$-fold degeneracies of 
the states in a level. Take for example the $\ell=(N-2)$ case already considered in (\ref{BIBONG1}) above. Substitution into (\ref{sim_trans}) 
amounts to:

\begin{eqnarray}
\Big( {\mathcal K}^2 +1 + V_{S^3}(\chi )\Big)\, \phi_{N(N -2)}(\chi) &=& 
\left( {\widetilde {\mathcal K}}^2+1 \right)\,\phi_{N  (N -2) }(\chi) -\frac{2b}{\cos\chi}\,{\mathbf F}^{-1}(\chi) \,\cos^{\ell}\chi 
\frac{ \partial P_1^{\alpha,\beta}(\sin \chi)}{
\partial \chi} \nonumber\\ 
 &=& 
(N-1 \big)^2(-b) \,{\widetilde S}_{(N-2)(N-2)}(\chi) \nonumber\\ 
 && + N^2\frac{(2N -1)}{4(N-1) }\,{\widetilde S}_{(N-1) (N-2)}(\chi) \nonumber\\ 
&& + (2N -1) (-b)\,{\widetilde S}_{(N-2)(N-2)}(\chi) \nonumber\\ 
&=& N^2 \,\phi_{N(N -2)}(\chi). 
\label{pre-space} 
\end{eqnarray} 

In the full space, obtained by multiplying (\ref{pre-space}) by $Y_{N -2}^m(\theta,\varphi)$ from the right,
the latter equation amounts to, 
\begin{eqnarray}
\left( {\widetilde {\mathcal K }}^2 +1 \right)\,\psi_{N(N -2)m}(\chi,\theta,\varphi) &-&
\frac{2b}{\cos\chi}{\mathbf F}^{-1}(\chi)\cos^{N -2} \chi 
\frac{ \partial P_1^{\alpha,\beta}(\sin \chi)}{\partial \chi}Y_{N-2}^m(\theta,\varphi)\nonumber\\ 
&=& N^2 \psi_{N(N-2)m}(\chi,\theta,\varphi).\nonumber\\
\label{That}
\end{eqnarray}
This simple exercise shows that the $N^2$-fold degeneracy in the spectrum of the trigonometric Scarf potential
with the $a$ parameter quantized to non-negative integer values can not be attributed in the usual way
to a geometric  $so(4)$ algebraic symmetry.
Yet, one is still left with the option of allowing for algebra invariants in the form of finite polynomials of
the transformed Casimir operator ${\widetilde {\mathcal K}}^2$ in (\ref{Cas_new_rep}) and (\ref{sim_trans1}).
If so, the r\'ole  of the gradiental term in ending up with $\phi_{N\ell }(\chi)$ functions belonging to
same  $N^2$ eigenvalue, despite their evident  decomposition into irreducible ${\widetilde {\mathcal K}}^2$
carrier space of different dimensionalities in (\ref{decomp}), could be  equally well played  by
properly chosen polynomial coefficients.
Take again as an example the $\ell =(N-2)$ case considered above.
The polynomial invariant of the transformed $so(4)$ algebra whose eigenvalue problem for $\phi_{N(N-2)}(\chi)$
is identical to that of ${\mathcal H}_{\mbox{\footnotesize Sc}}(\chi)$ on the same space, is given by

\begin{eqnarray}
\left[ \left({\widetilde {\mathcal K}}^2 +1\right)\frac{{\widetilde {\mathcal K}}^2+1 - (N-1)^2}{N^2-(N-1)^2}
 -b\left( {\widetilde {\mathcal K}}^2 +1 +(2N-1)\right)
\frac{{\widetilde {\mathcal K}}^2+1 - N^2}{(N-1)^2-N^2}\right]
{\Big(} {\widetilde Y}_{(N-1)(N-2)m}(\chi,\theta,\varphi)+&&\nonumber\\
{\widetilde Y}_{(N-2),(N-2),m}(\chi,\theta,\varphi){\Big)}
\equiv {\mathcal H}_{\mbox{\footnotesize Sc}}\psi_{N(N-2)m}(\chi,\theta,\varphi) 
=N^2\psi_{N(N-2)m}(\chi,\theta,\varphi).&&
\label{pol_1}
\end{eqnarray}   

Indeed, the term proportional to the projector \footnote{For the construction of such projectors see \cite{WuKi}. } 
on the $\widetilde{S}_{(N-2)(N-2)}(\chi)$ component in
(\ref{pre-space}), i.e. on the $(N-1)^2$ eigenvalue,
${\mathcal P}^{\left[(N-1)^2\right]}=\frac{{\widetilde {\mathcal K}}^2+1 - N^2}{(N-1)^2-N^2}$,
provides {\it by construction} same contribution to $(N-1)^2$,  namely,  $(2N-1)$, as the gradiental term, 
thus allowing for the factorization of  $N^2$ as a net eigenvalue.

In a similar way, the eigenvalue of $N^2$ to $\phi_{N (N-3)}(\chi)$, a space that contains according to the Table
the two lower dimensional components, ${\widetilde S}_{(N-2)(N-3)}(\chi)$, and ${\widetilde S}_{(N-3)(N-3)}(\chi)$,
can be understood as the following identity,

\begin{eqnarray}
{\Big(}
\left({\widetilde {\mathcal K}}^2 +1\right)
\left[\frac{{\widetilde {\mathcal K}}^2+1 - (N-1)^2}{N^2-(N-1)^2}\right]\,
\left[\frac{{\widetilde {\mathcal K}}^2+1 - (N-2)^2}{N^2-(N-2)^2}\right]&&\nonumber\\
-\frac{b}{2}\frac{(N-1)}{(N-2)}
\left({\widetilde  {\mathcal K}}^2+1 +(2N-1)\right)
\left[\frac{{\widetilde {\mathcal K}}^2+1 - N^2}{(N-1)^2-N^2}\right]\,
\left[\frac{{\widetilde {\mathcal K}}^2+1 - (N-2)^2}{(N-1)^2-(N-2)^2}\right]&&\nonumber\\
+\frac{b^2}{2}\left( {\widetilde {\mathcal K}}^2+1+4(N-1)\right)\left[\frac{{\widetilde {\mathcal K}}^2+1 - N^2}{(N-2)^2-N^2}\right]\,
\left[\frac{{\widetilde {\mathcal K}}^2+1 - (N-1)^2}{(N-2)^2-(N-1)^2}\right]{\Big)}&&\nonumber\\
\times \left( {\widetilde Y}_{(N-1)(N-3)m}(\chi,\theta,\varphi) + 
{\widetilde Y}_{(N-2)(N-3)m}(\chi,\theta,\varphi) + {\widetilde Y}_{(N-3)(N-3)m}(\chi,\theta,\varphi)\right)&&\nonumber\\
\equiv {\mathcal H}_{\mbox{\footnotesize Sc}}(\chi) \psi_{N(N-3)m}(\chi,\theta,\varphi))=N^2\psi_{N(N-3)m}(\chi,\theta,\varphi)).&&
\label{pol_2}
\end{eqnarray}   
In this case the polynomial is of third order in the Casimir operator ${\widetilde {\mathcal K}}^2$
in (\ref{sim_trans1}).
Along this line, appropriate  Casimir polynomials  for anyone of the degenerate states can be designed.
The above perturbative construction amounts to a potential specific diagonal matrix representation of the Scarf I Hamiltonian on $S^3$.\\

\noindent
Designing particle dynamics in terms of Hamiltonians as functions of certain Lie-algebra Casimir operators is
of common use and great utility in nuclear physics,
where the complexity of the systems presents an obstacle in the formulation of exactly soluble  potential problems.
This concept is of a fundamental importance within the powerful approach of the Interacting Boson Model (IBM) \cite{Iach}.
In the latter,  such descriptions of degeneracies are termed to as ``dynamics governed by generalized invariants of a symmetry algebra'', 
or, abbreviated, ``dynamical symmetries `` (not to be confused with the similar terminology used within the context
of algebra realizations in the full phase space, as is the case of the $H$-Atom and the Runge-Lentz vector \cite{Higgs}).
Conversely, the notion of a ``potential algebra'' confines to exactly solvable Hamiltonians which turn to 
be polynomials of first order in a standard  differential Casimir operator.  
In this fashion, and in reference to the IBM terminology,
the hydrogen-like degeneracy patterns in the spectrum under investigation can be understood  as a promotion of the 
geometric  $so(4)$ potential algebra of the $\sec^2$ interaction (corresponding to the free motion) to a dynamical 
$so(4)$ symmetry of the Scarf I potential (corresponding to the perturbed motion), with the algebra
being  in a representation unitarily nonequivalent to the hyper-spherical one.
Polynomials of the  type in (\ref{pol_1})-({\ref{pol_2}), when considered in terms of coefficients 
distributed at random around the Scarf I -specific values, would allow
for the description of $so(4)$ degeneracy patterns by means of wave functions fluctuating around 
the exact Scarf I solutions and could be of interest in simulation studies of systems with next-to Scarf I potential interactions.  
The scheme extends  to any $so(d+2)$.

\section{Conclusions and Perspectives}
The present study has been devoted to the explanation of the  emerging $so(4)$-type of degeneracy patterns in the
spectrum of the di-atomic (di-molecular)  trigonometric Scarf potential for 
the case in which  the parameter $a$ in (\ref{Scarfs_pot}) was allowed to take only non-negative  integer values, while $b$ remained  unrestricted.
We first  demonstrated that though perturbations by  $V_{S^3} (\chi) $ in (\ref{S3potential}) of the free motion 
are degeneracy conserving, they lead to the loss of the geometric $so(4)$ symmetry.  

Our argumentation was based on the observation that the wave functions of the perturbed motion behaved as  mixtures of functions, 
properly identified as genuine $so(4)$ representation functions in (\ref{new_repr_fu}),
and which, as illustrated by (\ref{BIBONG}), and the Table,
transformed as finite linear combinations of  $so(4)$ carrier spaces of different dimensionality.
Simultaneously, the decompositions presented themselves as finite power series expansions in the $b$ parameter, which permitted
to quantitatively keep track of the order to which the symmetry under discussion is gradually fading away.
Though the wave functions considered  constitute a $N^2$-fold degenerate multiplet, 
in not behaving as eigenstates to the standard  Casimir operator of the geometric $ so(4)$ under discussion,
prevents the interpretation of this symmetry as the culprit for the observed degeneracy,  unless the notion of an 
algebra invariant has not been generalized to an arbitrary function of the aforementioned Casimir operator.  
In such a case (\ref{pol_1})-(\ref{pol_2}), we showed that the diagonal matrix element of 
the Scarf I Hamiltonian for a given $\psi_{N\ell m}$ is  indistinguishable  from the 
$ \sum_{K=\ell}^{K=N-1}{\widetilde Y}_{K\ell m}\longrightarrow \psi_{N\ell m}$- 
transition matrix element of a polynomial of degree $n=(N-1-\ell) $ in the Casimir operator of the  
similarity transformed hyper-spherical algebra in eq.~(\ref{sim_trans1}).
In this way, we presented an explicit example on promoting by external scales a manifest  
geometric $so(d+2)$ symmetry of an exactly solvable potential
(the $so(4)$ of  $\sec^2$, in our case)
to an effective algebraic symmetry of the dynamics following a perturbation that retains the degeneracy.
Our findings are backed up by the established finite decompositions of the Jacobi polynomials, the key ingredients of the Scarf I solutions,  
in the basis of the  Gegenbauer polynomials, the key ingredients of the canonical $so(4)/so(d+2)$ representation functions. 

The present study points towards the possibility of non-standard $so(4)/so(2+d)$ realizations in the trigonometric Scarf potential problem,
a further option being deformations of the algebras as defined on the full phase space, where 
the momentum space for the sphere could be elaborated along the lines in ref.~\cite{KurtW}.
Alternatively, such a study could also be  worked out in the parametrization of the sphere via its
stereographic projection on an ambient linear space of one more dimension, with the aim to find the complete set of 
integrals of motion,  a scheme that has already been successfully  elaborated for the related Kepler-Coulomb problem 
on hyper-spheres of any dimension \cite{Hernaz} and 
which  leads  to algebra deformations.

To recapitulate, the goal of the present investigation has been to stress on the
possibility of retaining  degeneracy in the process of a perturbation.
The general interest in such an observation lies in the possibility to withdraw in a process of perturbation a 
fundamental geometric Lie algebraic 
symmetries by scales, such as temperatures, masses, lengths, without
leaving trace in the spectra. 
Such subtle symmetry losses remain undetectable at the level of the energy excitations but they have inevitably to show up in 
the disintegration modes of the system.

In view of the wide use of the hyper-spherical geometry in the description of many-body systems such as  Brownian motion \cite{Brown},
coherent states \cite{cohStates} etc. we expect our findings  reported here to acquire relevance. 
  

\end{document}